\documentclass[12pt, a4paper]{article}

\usepackage{graphicx}
\usepackage{comment}
\usepackage{simplemargins}
\usepackage{zed-csp}
\usepackage{IssamCh3}
\usepackage{mythesis}
\usepackage{datarefine}
\usepackage{jcst}
\usepackage{multicol}
\usepackage{setspace}
\usepackage{lineno,hyperref}

 \setleftmargin{2.6 cm}
 \setrightmargin{2.6 cm}
 \settopmargin{2.6 cm}
 \setbottommargin{2.6 cm}

\begin{document}

\input comment.sty

 \setcounter{page}{1}

 \setheadings {\rlap{\,}{\small\sl Archived 2019 \hfill J. Comput.
  Sci. \& Technol.\llap{{\,}}}}
  {\small\rlap{\,} {\small\sl Higher-Level H.W. Synthesis of The KASUMI Crypto.: \hfill}}
  {\small\hfill  {\small\sl  J. Comput. Sci. \& Technol.}   \llap{\,}}
 \def\footnote{}

\noindent{\Large\bf The link to the formal publication is via\\ {\small \url{https://doi.org/10.1007/s11390-007-9007-9}}\\ Higher-Level Hardware Synthesis of The KASUMI
Algorithm\vskip 2ex \noindent \normalsize\rm Issam W. Damaj

\vskip 1ex \noindent \small{\it Electrical and Computer Engineering Department, Hariri
Canadian Academy of Sciences and Technology, Meshref P.O.Box: 10 Damour- Chouf 2010 Lebanon}\\[1mm]

\noindent E-mail: damajiw@hariricanadian.edu.lb\\[1mm]


{\noindent {\small\bf Abstract} \quad \small Programmable Logic
Devices (\textit{PLDs}) continue to grow in size and currently
contain several millions of gates. At the same time, research
effort is going into higher-level hardware synthesis methodologies
for reconfigurable computing that can exploit \textit{PLD}
technology. In this paper, we explore the effectiveness and extend
one such formal methodology in the design of massively parallel
algorithms. We take a step-wise refinement approach to the
development of correct reconfigurable hardware circuits from
formal specifications. A functional programming notation is used
for specifying algorithms and for reasoning about them. The
specifications are realised through the use of a combination of
function decomposition strategies, data refinement techniques, and
off-the-shelf refinements based upon higher-order functions. The
off-the-shelf refinements are inspired by the operators of
Communicating Sequential Processes (\textit{CSP}) and map easily
to programs in \textit{Handel-C} (a hardware description
language). The \textit{Handel-C} descriptions are directly
compiled into reconfigurable hardware. The practical realisation
of this methodology is evidenced by a case studying the third
generation mobile communication security algorithms. The
investigated algorithm is the \textit{KASUMI} block cipher. In
this paper, we obtain several hardware implementations with
different performance characteristics by applying different
refinements to the algorithm. The developed designs are compiled
and tested under \textit{Celoxica's} \textit{RC-1000}
reconfigurable computer with its 2 million gates \textit{Virtex-E}
\textit{FPGA}. Performance analysis and evaluation of these
implementations are included.


\begin{multicols}{2}

\section{\normalsize\bf\quad Introduction}

\quad\ The rapid progress and advancement in electronic chips technology provides a variety of new
implementation options for system engineers. The choice varies between the flexible programs
running on a general purpose processor \textit{(GPP)} and the fixed hardware implementation using
an application specific integrated circuit (\textit{ASIC}). Many other implementation options
present, for instance, a system with a \textit{RISC} processor and a \textit{DSP} core. Other
options include graphics processors and microcontrollers. Specialist processors certainly improve
performance over general-purpose ones, but this comes as a quid pro quo for flexibility. Combining
the flexibility of \textit{GPPs} and the high performance of \textit{ASICs} leads to the
introduction of reconfigurable computing (\textit{RC}) as a new implementation option with a
balance between versatility and speed.

Field Programmable Gate Arrays (\textit{FPGAs}), nowadays are
important components of \textit{RC}-systems, have shown a dramatic
increase in their density over the last few years. For example,
companies like \textit{Xilinx} \cite{1Ka} and \textit{Altera}
\cite{127} have enabled the production of \textit{FPGAs} with
several millions of gates, such as in \textit{Virtex-II Pro} and
\textit{Stratix-II} \textit{FPGAs}. The versatility of
\textit{FPGAs}, opened up completely new avenues in
high-performance computing.


The traditional implementation of a function on an \textit{FPGA}
is done using logic synthesis based on \textit{VHDL}, Verilog or a
similar \textit{HDL} (hardware description langauge). These
discrete event simulation languages are rather different from
languages, such as \textit{C}, \textit{C++} or \textit{JAVA}. An
interesting step towards more success in hardware compilation is
to grant a higher-level of abstraction from the point of view of
programmer. Designer productivity can be improved and
time-to-market can be reduces by making hardware design more like
programming in a high-level langauge. Recently, vendors have
initiated the use of high-level languages dependent tools like
\textit{Handel-C} \cite{129}, \textit{Forge} \cite{130},
\textit{Nimble} \cite{131}, and \textit{SystemC} \cite{133}.


With the availability of powerful high-level tools accompanying
the emergence of multi-million \textit{FPGA} chips, more emphasis
should be placed on affording an even higher level of abstraction
in programming reconfigurable hardware. Building on these research
motivations, in the work in hand, we extend and examine a
methodology whose main objective is to allow for a higher-level
correct synthesis of massively parallel algorithms and to map
(compile) them onto reconfigurable hardware. Our main concern is
with behavioural refinement, in particular the derivation of
parallel algorithms. The presented methodology systematically
transforms functional specifications of algorithms into parallel
hardware implementations. It builds on the work of Abdallah and
Hawkins \cite{6,140b} extending their treatment of data and
process refinement.

This paper is divided so that some of the following sections
introduce the adopted development methodology. Section~\ref{bkgnd}
presents the theoretical background. In Section~\ref{App}, we put
some emphasis on the approach to develop different implementations
of the KASUMI cryptographic algorithm. The following section
details the development steps. Section~\ref{RHI} demonstrates
selected implementations. In Section~\ref{PAE}, we analyze and
evaluate the performance of the suggested implementations.
Finally, Section~\ref{Con} concludes the paper.

\section{\normalsize\bf\quad The Development Method}

The suggested development model adopts the transformational
programming approach for deriving massively parallel algorithms
from functional specifications (See Figure~\ref{RDM}). The
functional notation is used for specifying algorithms and for
reasoning about them. This is usually done by carefully combining
a small number of higher-order functions that serve as the basic
building blocks for writing high-level programs. The systematic
methods for massive parallelisation of algorithms work by
carefully composing an "off-the-shelf" massively parallel
implementation of each of the building blocks involved in the
algorithm. The underlying parallelisation techniques are based on
both pipelining and data parallelism.

\begin{figure*}
  \begin{center}
    \includegraphics [scale= 0.8]
                      {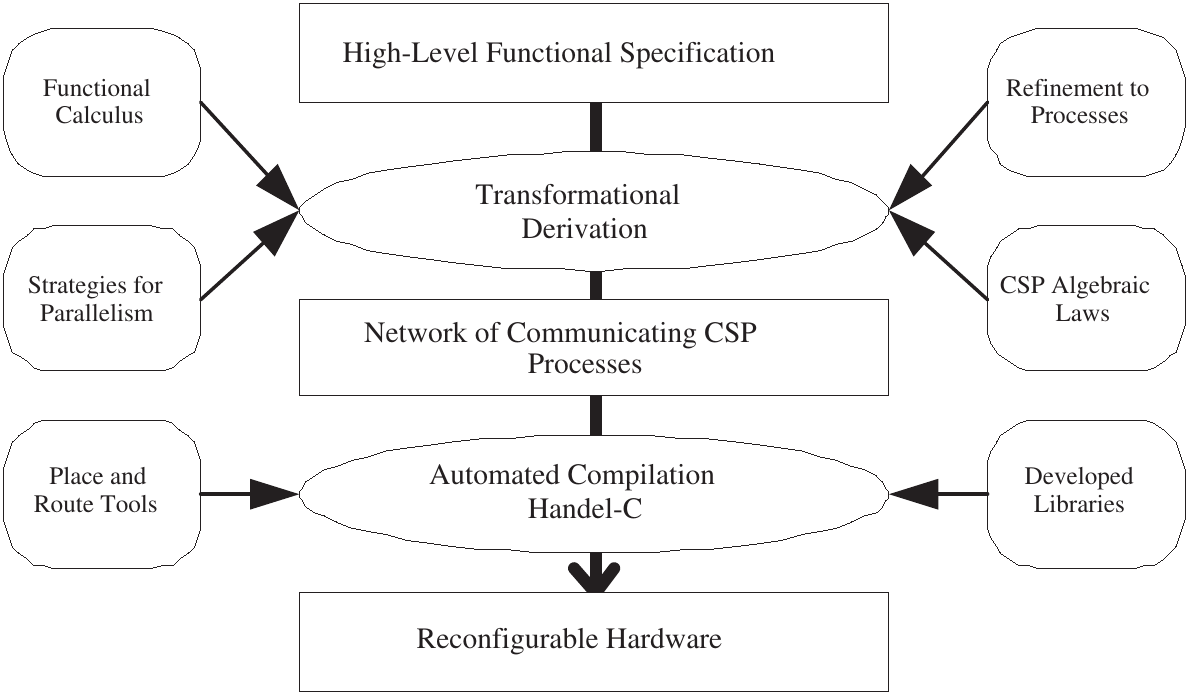}
    \caption{An overview of the transformational derivation and the hardware realisation processes.}
    \label{RDM}
  \end{center}
\end{figure*}

Higher-order functions, such as \textit{map}, \textit{filter}, and
\textit{fold}, provide a high degree of abstraction in functional
programs \cite{11}. Not only they do allow clear and succinct
specifications for a large class of algorithms, but they also are
ideal starting points for generating efficient implementations by
a process of mathematical calculation using Bird-Meertens
Formalism (\textit{BMF}). The essence of this approach is to
design a generic solution once, and to use instances of the design
many times for various applications. Accordingly, this approach
allows portability by implementing the design on different
parallel architectures.

In order to develop generic solutions for general parallel
architectures it is necessary to formulate the design within a
concurrency framework such as Hoare's\textit{CSP} \cite{16}. Often
parallel functional programs show peculiar behaviours which are
only understandable in the terms of concurrency rather than
relying on hidden implementation details. The formalisation in
\textit{CSP} (of the parallel behaviour) leads to better
understanding and allows for analysis of performance issues. The
establishment of refinement concepts between functional and
concurrent behaviours may allow systematic generation of parallel
implementations for various architectures.

The previous stages of development require a back-end stage for realising the developed designs.
We note at this point that the \textit{Handel-C} language relies on the parallel constructs in
\textit{CSP} to model concurrent hardware resources. Mostly, algorithms described with
\textit{CSP} could be implemented with \textit{Handel-C}. Accordingly, this langauge is suggested
as the final reconfigurable hardware realisation stage in the proposed methodology. It is noted
that, for the desired hardware realisation, \textit{Handel-C} enables the integration with
\textit{VHDL} and \textit{EDIF} (Electronic Design Interchange Format) and thus various synthesis
and place-and-route tools.


\section{\normalsize\bf\quad\  Background} \label{bkgnd}

\quad\ Abdallah and Hawkins defined in \cite{140b} some constructs
used in the development model. Their investigation looked in some
depth at data refinement; which is the means of expressing
structures in the specification as communication behaviour in the
implementation.

\subsection{\normalsize\bf\quad\ Data Refinement}

\quad\ In the following we present some datatypes used for refinement, these are stream, vector,
and combined forms.

The stream is a purely sequential method of communicating a group
of values. It comprises a sequence of messages on a channel, with
each message representing a value. Values are communicated one
after the other. Assuming the stream is finite, after the last
value has been communicated, the end of transmission
(\textit{EOT}) on a different channel will be signaled. Given some
type \textit{A}, a stream containing values of type \textit{A} is
denoted as $\langle A \rangle$.

Each item to be communicated by the vector will be dealt with
independently in parallel. A vector refinement of a simple list of
items will communicate the entire structure in a single. Given
some type \textit{A}, a vector of length \textit{n}, containing
values of type \textit{A}, is denoted as $\lfloor A \rfloor_{n}$.

Whenever dealing with multi-dimensional data structures, for example, lists of lists,
implementation options arise from differing compositions of our primitive data refinements -
streams and vectors. Examples of the combined forms are the Stream of Streams, Streams of Vectors,
Vectors of streams, and Vectors of Vectors. These forms are denoted by: $\langle
S_{1},S_{2},...,S_{n}\rangle$ , $\langle V_{1},V_{2},...,V_{n}\rangle$, $\lfloor
S_{1},S_{2},...,S_{n}\rfloor$, and $\lfloor V_{1},V_{2},...,V_{n}\rfloor$.

\subsection{\normalsize\bf\quad\ Process Refinement}

\quad\ The refinement of the formally specified functions to
processes is the key step towards understanding possible parallel
behaviour of an implementation. In this section, the interest is
in presenting refinements of a subset of functions - some of which
are higher-order. A bigger refined set of these functions is
discussed in \cite{6}.

Generally, These highly reusable building blocks can be refined to
\textit{CSP} in different ways. This depends on the setting in
which these functions are used (i.e. with streams, vectors etc.),
and leads to implementations with different degrees of
parallelism. Note that we don't use \textit{CSP} in a totally
formal way, but we use it in a way that facilitates the
\textit{Handel-C} coding stage later. Recall for the following
subsections that values are communicated through as an
\textit{elements} channel, while a single bit is communicated
through another \textit{eotChannel} channel to signal the end of
transmission (\textit{EOT}).

\subsubsection{\normalsize\bf\quad\ Basic Definitions}

\quad\ The produce/store process (\textit{PRD}/\textit{STORE}) is
fundamental to process refinement. It is used to produce/store
values on/from the channels of a certain communication construct
(\textit{Item}, \textit{Stream}, \textit{Vector}, and so on).
These values are to be received and manipulated by another
processes.

\quad\ The feed operator in \textit{CSP} models function
application. The feed operator is written $\rhd$.

{\singlespace
\[
P \rhd Q = \\ (P [mid / out] ~ || ~ Q [mid / in]) \verb"\" \{ mid \}
\]
}


\quad\ Consider a potential refinement for $f$, a process $F$. The
operator $\sqsubseteq$ denotes a process refinement, where the
left hand side is a function, and the right hand side is a
process. To state that $f$ is refined to $F$, or in other words,
the process $F$ is a valid refinement of the function $f$, the
following may be used:

\[
f \sqsubseteq F
\]

These rules were proven once \cite{6}, and in this paper we use
them systematically to refine the functional specification into a
network of communicating processes.


\subsubsection{\normalsize\bf\quad\ Process Refinement of Higher-order Functions }

\quad\ Now the attention is turned to the refinement of
higher-order functions presented in \cite{140b}, showing the
refinement of the high-order function \textit{map} as an instance.
Employing this function in stream and vector settings is
presented.

\paragraph{\normalsize\bf\quad\ Streams}

\quad\ A process implementing the functionality of $map ~ f$ in
stream terms should input a stream of values, and output a stream
of values with the function $f$ applied.

In general, the handling of the \textit{EOT} channels will be the same. However, the handling of
the value will vary depending on the type of the elements of the input and output stream.

\[
SMAP(F) =\\
\mu  X ~ \bullet  in.eotChannel ~ ? ~ eot \rightarrow \\ out.eotChannel ~ ! ~ eot \rightarrow SKIP \\
                   \choice \\
                   F [in.elements.channel / in, \\ out.elements.channel / out] ; X
\]

\paragraph{\normalsize\bf\quad\ Vectors}

\quad\ In functional terms, the functionality of $map ~ f$ in a
list setting is modelled by $vmap ~ f$ in the vector setting.
Consider $F$ as a valid refinement of the function $f$. The
implementation of $VMAP$ can then proceed by composing $n$
instances of $F$ in parallel, and directing an item from the input
vector to each instance for processing. In \textit{CSP} we have:

{\singlespace
\begin{equ}
VMAP_{n}(F) & = & \interleave_{i=1}^{i=n} \\ F [ in_{i} / in, out_{i} / out]
\end{equ}
}

\subsection{\normalsize\bf\quad\ Handel-C as a Stage in the Development Model}

\quad\ Based on datatype refinement and the skeleton afforded by process refinement, the desired
reconfigurable circuits are built. Circuit realisation is done using \textit{Handel-C}, as it is
based on the theories of \textit{CSP} \cite{16} and \textit{Occam} \cite{B3}.

From a practical standpoint, each refined datatype is defined as a structure in \textit{Handel-C},
while each process is implemented as a \textit{macro} \textit{procedure}. We divide the constructs
corresponding to the \textit{CSP} stage into 2 main categories for organisation purposes. The
first category represents the definitions of the refined datatypes. The second category implements
the refined processes.

The refined processes are divided into different groups; the
\textit{utility}, \textit{basic}, \textit{higher-order} processes.
A separate group contains the macros that handle the \textit{FPGA}
card setup and general functionality.

The datatypes definitions are implemented using structures. This
method supports recursive as well as simple types. The definition
for an \textit{Item} of a type \textit{Msgtype} is a structure
that contains a communicating channel of that type.

{\singlespace  {\small
\begin{verbatim}
 #define Item(Name, Msgtype)
    struct {
        chan Msgtype    channel;
        Msgtype         message;
        } Name
\end{verbatim}}}

For generality in implementing processes the type of the communicating structure is to be
determined at compile time. This is done using the \textit{typeof} type operator, which allows the
type of an object to be determined at compile time. For this reason, in each structure we declare
a \textit{message} variable of type \textit{Msgtype}.

A stream of items, called \textit{StreamOfItems}, is a structure
with three declarations a communicating channel, an \textit{EOT}
channel, and a \textit{message} variable \cite{140b}:

{\singlespace  {\small
\begin{verbatim}
 #define StreamOfItems(Name, Msgtype)
    struct {
        Msgtype         message;
        chan Msgtype    channel;
        chan Bool       eotChannel;
        } Name
\end{verbatim}}}

A vector of items, called \textit{VectorOfItems}, is a structure
with a variable \textit{message} and another array of
sub-structure elements \cite{140b}.

{\singlespace  {\small
\begin{verbatim}
 #define VectorOfItems(Name, n, Msgtype)
    struct {
        struct {
            chan Msgtype    channel;
            } elements[n];
        Msgtype     message;
        } Name
\end{verbatim}}}

Other definitions are possible, but it affects the way a channel
is called using the structure member operator (.).


The utility processes used in the implementation are related to
the employed datatypes. The \textit{Handel-C} implementation of
these processes relies on their corresponding \textit{CSP}
implementation. In the following, we present an instance of these
utility macros.

{\singlespace  {\small
\begin{verbatim}
 macro proc ProduceItem(Item, x){
    Item.channel ! x;}

 macro proc StoreItem(Item, x){
    Item.channel ? x;}
\end{verbatim}}}


This group of macros represents the fine-grained processes. A
sample basic macro procedure \textit{Addition} is included as an
example.

{\singlespace  {\small
\begin{verbatim}
 macro proc
  Addition(xItem, yItem, output){
   typeof (xItem.message) x,y;
   xItem.channel ? x;
   yItem.channel ? y;
   output.channel ! (x + y);}
\end{verbatim}}}

\subsubsection{\normalsize\bf\quad\ Higher-Order Processes Macros}

\quad\ An example for an implementation in \textit{Handel-C} of
the CSP refinement of a higher-order function (\textit{map}) in
its vector setting is done as follows:

{\singlespace  {\small
\begin{verbatim}
 macro proc
  VMAP (n, vectorin, vectorout, F) {
   typeof (n) c;
   par (c = 0 ; c < n ; c++){
      F(vectorin.elements[c],
        vectorout.elements[c]);}}
\end{verbatim}}}

In a similar procedure to what have been introduced before, the
implementations of the stream and vector settings
\textit{SZipWith} and \textit{VZipWith} are straightforward.


\quad\ Different tools are used to measure the performance metrics
used for the analysis. These tools include the design suite
(\textit{DK}) from \textit{Celoxica}, where we get the number of
\textit{NAND} gates for the design as compiled to the Electronic
Design Interchange Format (\textit{EDIF}). The \textit{DK} also
affords the number of cycles taken by a design using its
simulator. Accordingly, the speed of a design could be calculated
depending on the expected maximum frequency of the design. The
maximum frequency could be determined by the timing analyzer. To
get the practical execution time as observed from the computer
hosting the RC-1000, the \textit{C++} high-precision performance
counter is used. The information about the hardware area occupied
by a design, i.e. number of Slices used after placing and routing
the compiled code, is determined by the \textit{ISE} place and
route tool from \textit{Xilinx}.


\section{\normalsize\bf\quad\ The Third Generation of Mobile System Security
Algorithms}\label{App}

\quad\ The \textit{KASUMI} is a modern and strong encryption
algorithm designed for the use in the Third Generation Partnership
Project (\textit{3GPP}) security functions for mobile systems
\cite{k3}. \textit{KASUMI} ciphers a 64-bit input data block by
repeating a round procedure 8 times. The round composes a 32-bit
non-linear mixing block (\textit{FO}) and a 32-bit linear mixing
block (\textit{FL}). The \textit{FO}-block is an iterated
"ladder-design" consisting of 3 rounds of a 16-bit non-linear
mixing block \textit{FI}. In turn, \textit{FI} randomising
function is defined as a 4-round structure using non-linear
look-up tables \textit{S7} and \textit{S9}. All functions involved
will mix the data input with key. The used \textit{S7} and
\textit{S9} have been designed in a way that avoids linear
structures in \textit{FI} - this fact has been confirmed by
statistical testing. Each functional component of \textit{KASUMI}
has been carefully studied to reveal any weakness that could be
used as a basis for an attack on the entire algorithm.  The fact
that the key schedule of \textit{KASUMI} is very simple did not
constitute any real weakness. There seems to be no gain in
practice by making it more complicated.

Hardware implementation of this cryptographic algorithm is
currently an active area of research. The \textit{KASUMI} was
addressed by HoWon et al \cite{k7}, and Alcantara et al \cite{k9}.
\textit{Intel} \cite{139} proposed architecture processors for
\textit{3G} control including the \textit{KASUMI}. Moreover,
\textit{SCIWORX} \cite{k14} produced a system board for the
\textit{KASUMI} cipher.


\section{\normalsize\bf\quad\ Formal Functional Specification} \label{FFS}

\quad\ We will consider the following specifications for the key
scheduler, and the main algorithm (\textit{KASUMI}). The key
scheduler takes the private key as an input, and outputs a desired
set of subkeys. This set of subkeys is of 4 packs (See
Figure~\ref{KeySchGen}). The \textit{KASUMI} takes two inputs, the
generated subkeys and the input data, and it gives their
corresponding output.

\begin{figure*}
  \begin{center}
    \includegraphics [scale=0.9]
                      {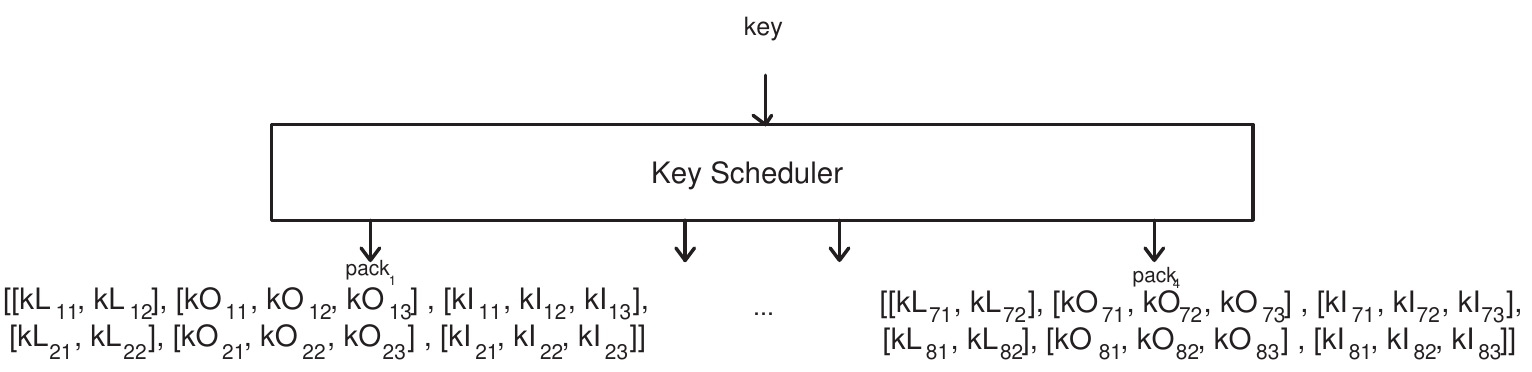}
    \caption{Key scheduling building blocks}
    \label{KeySchGen}
  \end{center}
\end{figure*}

Generally, the functional specification style applied throughout this research uses higher-order
functions as the main keys for later parallelism. As a start, we define some types to be used in
the following formal specification:

{\singlespace {\small \begin{verbatim}
 type Private = [Bool]
 type SubKey = [Bool]
 type DataBlock = [Bool]
\end{verbatim}}}

The following specifications are also tested using the \textit{Hugs98} \textit{Haskell} compiler.


\subsection{\normalsize\bf\quad\ Key Scheduling}

\quad\ As shown in Figure~\ref{KeySchGen}, the 64 16-bit subkeys are organised into 4 packs of 8
sets of subkeys $kL_{i1},$ $kL_{i2},$ $kO_{i1},$ $kO_{i2},$ $kO_{i3},$ $kI_{i1},$ $kI_{i2},$ and
$kI_{i3}$, where \textit{i} is an index corresponding to the round number where a subkey is to be
used. These subkeys are generated from the 128-bit encryption private key.

Key scheduling is specified as the function \textit{keySchedule} that inputs a private key and
outputs 4 packs of subkeys. We divide each pack into 6 groups for later ease of distribution to
the encrypting rounds. Each group is a list of subkeys selected from the predefined lists
$kL_{i1}, kL_{i2}, kO_{i1}, kO_{i2}, kO_{i3}, kI_{i1}, kI_{i2},$ and $kI_{i3}$. For instance, the
first pack would contain:
\\

{\singlespace {\small $[[kL_{11}, kL_{12}], [kO_{11}, kO_{12}, kO_{13}], [kI_{11}, kI_{12},
kI_{13}],$ $[kL_{21}, kL_{22}], [kO_{21}, kO_{22}, kO_{23}], [kI_{21}, kI_{22}, kI_{23}]]$}}
\\

The specification of \textit{keySchedule} is formalised as follows:

{\singlespace {\small \begin{verbatim}
 keySchedule :: Private -> [[[Subkey]]]
 keySchedule key = merge(g)
   where
   [kLi1, kOi1, kOi2, kOi3]  =
   mapWith
    (map map [(shift 1), (shift 5),
              (shift 8), (shift 13)])
    (mapWith [id, (shift 1),
              (shift 5),(shift 6)]
    (copy (segs 16 key) 4))

   [kLi2, kIi1, kIi2, kIi3]  =
   mapWith [(shift 2), (shift 4),
            (shift 3), (shift 7)] (
           copy ks' 4)

   ks' = zipWith  fullexor (segs 16 key)
   (map itob [291, 17767, 35243, 52719,
              65244, 47768, 30292, 12816])

   g = ((map group).transpose)
       [kLi1, kLi2, kOi1, kOi2, kOi3,
       kIi1, kIi2, kIi3]

   merge(gr) =
    [(gr!!i)++(gr!!(i+1)) |
                        i <- [0,2,4,6]]
\end{verbatim}}}

The function \textit{keySchedule} generates the subkeys by firstly determining the predefined
\textit{ks} and \textit{ks'}. \textit{ks} is specified using the function \textit{segs} as
\textit{(segs 16 key)}. Recall that \textit{segs} selects \textit{n} sublists from a list
\textit{xs}.

After specifying \textit{ks}, we formalise the computation for \textit{ks'} using the higher-order
function \textit{zipWith} zipping two lists with the function \textit{exor}. These lists
corresponds to \textit{ks} and \textit{C}. After \textit{ks} and \textit{ks'} are ready,
\textit{KASUMI} subkeys are determined employing the higher-order functions \textit{mapWith} and
\textit{map}. Also, using the functions \textit{shift} and \textit{copy}.

Finally, the functions \textit{group} and \textit{transpose} arrange the subkeys in the form
mentioned earlier. The arranged groups are then merged into final 4 packs. To easily understand
these steps we include the chart shown in Figure~\ref{keySchChart}.



\subsection{\normalsize\bf\quad\ The KASUMI Block Cipher}

\quad\ The \textit{KASUMI} block cipher has two inputs, a 64-bit data block in addition to the
private key. The corresponding ciphered output is also a 64-bit data block. In this specification,
we suggest the division of the \textit{KASUMI} structure into 4 similar rounds. Where each single
round is of two subrounds, called first and second subrounds. The 4 generated packs of subkeys
(using the function \textit{keySchedule}) are distributed to the \textit{KASUMI} 4 rounds
respectively. The total 8 subrounds of the \textit{KASUMI} constitute a Feistel network. This is
visualised in Figure~\ref{KASUMIBlock}.

\begin{figure}
  \begin{center}
    \includegraphics [scale=0.5]
                      {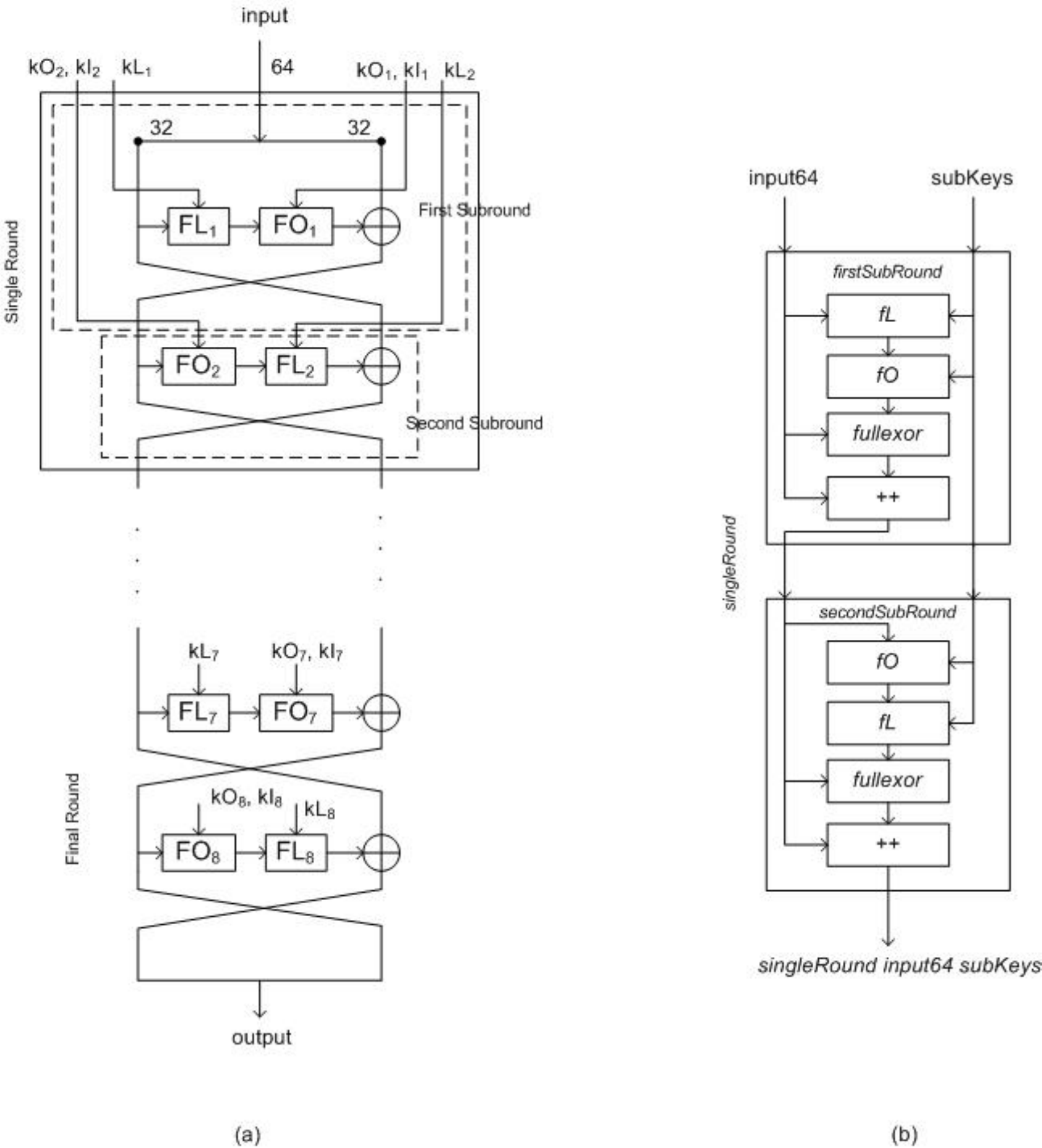}
    \caption{(a) The KASUMI block. (b) A single round}
    \label{KASUMIBlock}
  \end{center}
\end{figure}

\textit{KASUMI} is formally specified as the function \textit{kasumi} which inputs two lists of
\textit{bool} \textit{input} and \textit{key}. This function outputs a list of \textit{bool}
corresponding to the ciphered data. The specification is done by folding a function
\textit{singleRound} with the input over the generated subkeys packs. With respect to the network
shape, the foldable single round is specified as the function \textit{singleRound}.

{\singlespace {\small \begin{verbatim}
 kasumi ::
   DataBlock -> Private -> DataBlock
 kasumi input key =
   foldl singleRound input
                     (keyScheduling key)
\end{verbatim}}}

A single round is of two blocks, the odd block formalised as the function \textit{firstSubRound}
and the even round formalised as the function \textit{secondSubRound}. The function
\textit{singleRound} is specified as the functional composition of the functions
\textit{firstSubRound} and \textit{secondSubRound}. The inputs to the function
\textit{singleRound} are an input block of data and a single pack of subkeys.

{\singlespace {\small \begin{verbatim}

 singleRound ::
 DataBlock -> [[Subkeys]] -> DataBlock

 singleRound input64 subKeys =
 secondSubRound (firstSubRound input64
                               subKeys)
\end{verbatim}}}

The function \textit{firstSubRound} could be described as follows. It firstly takes the 64-bit
data input block and divides it into two left and right 32-bit words as shown in
Figure~\ref{KASUMIBlock}. It also inputs a pack of subkeys and distributes them to their specific
destinations. The data input left half is passed to a function \textit{fL}, which corresponds to
the \textit{FL} block. The function \textit{fL} forwards its output to a function \textit{fO} (the
functional specification of the \textit{FO} block). The output from the function \textit{fO} is
XORed with the right half of the input data giving the final left half \textit{l1}. The
\textit{firstSubRound} outputs a 64-bit word, which is the concatenation of the final left half
with the initial left half. Also, it outputs the subkeys needed for the second subround.

{\singlespace {\small \begin{verbatim}
 firstSubRound ::
 DataBlock -> [[SubKey]] ->
            (DataBlock,[[SubKey]])

 firstSubRound
  input64 [kLo, kOo, kIo, kLe, kOe, kIe] =
   (l1++r1, [kLe, kOe, kIe])

  where
  [r1, r0]=
    [take 32 input64, drop 32 input64]

  [l1, t1, t2]=
    [(fullexor r0 t2),
     (fL r1 kLo),
     (fO t1 kOo kIo)]
\end{verbatim}}}

The function \textit{secondSubRound} divides the input 64-bit data block into two left and right
halves. The left half with the suitable subkeys are passed to the function \textit{fO}. The output
from the function \textit{fO} is forwarded to the function \textit{fL}. The output from the
function \textit{fL} is XORed with the input right half to give the final left half \textit{l2}.
The \textit{secondSubRound} outputs a 64-bit word, which is the concatenation of the final left
half with the final right half \textit{r2}.

{\singlespace {\small \begin{verbatim}
 secondSubRound ::
  (DataBlock, [[SubKey]]) -> DataBlock

 secondSubRound
  (input64, [kL, kO, kI])  = l2 ++ r2

  where
  [r1, r2]= [drop 32 input64,
             take 32 input64]

  [l2, t1, t2]= [(fullexor r1 t2),
                 (fO r2 kO kI),
                 (fL t1 kL)]
\end{verbatim}}}

The remaining \textit{fL}, \textit{fI}, \textit{fO}, \textit{s7},
and \textit{s9} building blocks are specified in a similar style.


\section{\normalsize\bf\quad\ Algorithms Refinements} \label{AF}

\quad\ We move now to the second stage of development following
the same proposed method. The refinement of the key scheduling,
and the \textit{KASUMI} specifications are presented in the
following subsections.

\subsection{\normalsize\bf\quad\ Key Scheduling}

\quad\ Getting closer to hardware implementation, the general datatypes used in specifying the
function \textit{keySchedule} are refined as follows:
\\

$keySchedule :: Int128 \rightarrow \lfloor\lfloor\lfloor Int16 \rfloor\rfloor_{6}\rfloor_{4}$
\\

The key is a 128-bit Integer item, and the output packs of groups of lists can be refined to a
vector of 4 vectors, each of 6 vectors of 16-bit Integer items. The refined processes
\textit{KEYSCHEDULE} corresponds to the function \textit{keySchedule}.
\\

$keySchedule \sqsubseteq KEYSCHEDULE$
\\

From the specification, the process \textit{KEYSCHEDULE} inputs the key and then it divides it
into segments using the process \textit{SEGS} the refinement of \textit{segs}. These segments are
broadcasted to be later used for 5 times. At this point, two parallel events could occur
corresponding to the right and left branches depicted in Figure~\ref{keySchP}. The right branch of
processes refines the following part of the specification:

{\singlespace {\small \begin{verbatim}
 ks' = zipWith  fullexor (segs 16 key)
      (map itob [291, 17767, 35243,
                 52719, 65244, 47768,
                 30292, 12816])

 [kLi2, kIi1, kIi2, kIi3]  =
   mapWith [(shift 2), (shift 4),
            (shift 3), (shift 7)]

            (copy ks' 4)
\end{verbatim}}}

To compute for \textit{ks'} the vector setting refinement of \textit{zipWith} (\textit{VZIPWITH})
is used. Then the vector refinement of \textit{mapWith}, \textit{VMAPWITH}, is used to compute for
the first set of subkeys.

The parallel left branch of processes computes for the second set of subkeys by piping two
instances of the refined process \textit{VMAPWITH}. This refines the following recalled
specification:

{\singlespace {\small \begin{verbatim}
 [kLi1, kOi1, kOi2, kOi3]  =
 mapWith (map map
  [(shift 1),(shift 5),
  (shift 8), (shift 13)])

  (mapWith
  [id, (shift 1), (shift 5),(shift 6)]
   (copy (segs 16 key) 4))
\end{verbatim}}}

The remaining processes are used to refine the functions responsible for ordering the subkeys in
the suggested form - packs of groups of lists. The complete network of processes (see
Figure~\ref{keySchP}) is described as follows:
\\

{\singlespace

\noindent $KEYSCHEDULE = (32\rhd SEGS) \parallel IBROADCAST_{5}[d/out] \parallel$

\noindent$($

 \  $(([291, 17767, 35243, 52719, 65244,$

 \ $ 47768, 30292, 12816]\rhd VZIPWITH(EXOR))$

 \ $\gg_{8} IBROADCAST_{4}[d/out] \parallel$

 \ $VMAPWITH([SHIFT(2), SHIFT(4),$

 \ $SHIFT(3), SHIFT(7)])$

 \ $)$

 \ $\parallel$

 \ $($

 \ $VMAPWITH([SHIFT(1), SHIFT(1),$

 \ $SHIFT(5), SHIFT(6)])\gg_{4}$

 \ $VMAPWITH[ID, VMAP(SHIFTL(5)),$

 \ $VMAP(SHIFTL(8)), VMAP(SHIFTL(13))]$

 \ $)$

\noindent $)\gg_{8} TRANSPOSE \gg_{8} VMAP(GROUP) \gg_{8} MERGE$
\\

\noindent where
\\

$group \sqsubseteq GROUP$

$merge \sqsubseteq MERGE$

$shift \sqsubseteq SHIFTL$ }
\\

The process \textit{TRANSPOSE} is the standard matrix transpose.

\begin{figure*}[htpb]
  \begin{center}
    \includegraphics [scale=0.8]
                      {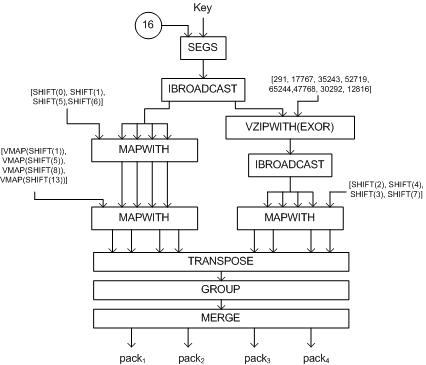}
    \caption{The process KEYSCHEDULE}
    \label{keySchP}
  \end{center}
\end{figure*}


\subsection{\normalsize\bf\quad\ The KASUMI Block Cipher}

\quad\ The \textit{KASUMI} block is the main ciphering part used for the confidentiality and
integrity algorithms standardised for \textit{3GPP}. Based on the functional specification stage
of development, we suggest two refined designs for implementing the \textit{KASUMI} block. The
first is a 4 rounds pipelined design, while the second proposes a single round stream-based
design.

\subsubsection{\normalsize\bf\quad\ First Design}

\quad\ In this design, we construct a fully pipelined network implementing the \textit{KASUMI}
block. Four single rounds are replicated to work in parallel forming a pipeline of processes.
Accordingly, this design is expected to have a high degree of parallelism, and therefore to be
highly efficient. However, this processes-replicating implementation will require the use of large
amounts of processing resources.

The first step in refining the function \textit{kasumi} observes its inputs as items with a
precision of 64 bits for the data block and 128 bits for the key. This is described as follows:
\\

$kasumi :: Int64 \rightarrow Int128 \rightarrow Int64$

\noindent where $kasumi \sqsubseteq KASUMI$
\\

As for this design, the four groups of subkeys are piped from the process \textit{KEYSCHEDULE} to
the replicated \textit{SINGLEROUND} processes. The \textit{foldl} higher-order function in this
case is refined to its vector setting \textit{VVFOLDL}. Thus, the process \textit{KASUMI} is
refined as follows:
\\

$KASUMI = KEYSCHEDULE \parallel VVFOLDL(SINGLEROUND)$
\\

Note that the upper input to each \textit{SINGLEROUND} is a list of list of subkeys, refined as a
vector of vectors. This is depicted in Figure~\ref{KASUMI1D}.

\begin{figure*}
  \begin{center}
    \includegraphics [scale=0.8]
                      {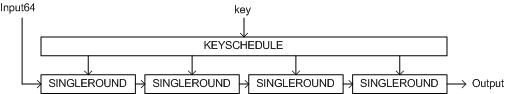}
    \caption{The process KASUMI, first fully-pipelined design}
    \label{KASUMI1D}
  \end{center}
\end{figure*}

Moving to the refinement \textit{KASUMI} sub-blocks, datatypes employed in the function
\textit{singleRound} could be refined as follows:
\\

$singleRound :: Int64 \rightarrow \lfloor [Int16] \rfloor_{6} \rightarrow Int64$
\\

\noindent where $singleRound \sqsubseteq SINGLEROUND$

Recall the functional specification for a \textit{singleRound}, we have:

{\singlespace {\small
\begin{verbatim}
 singleRound input64 subKeys =
    secondSubRound
    (firstSubRound input64 subKeys)
\end{verbatim}}}

This functional composition is refined to piping of two processes
\textit{FIRSTSUBROUND} and \textit{SECONDSUBROUND}. The process
\textit{SINGLEROUND} is depicted in Figure~\ref{figs} (a) and
described as follows:
\\

$SINGLEROUND = FIRSTSUBROUND \gg SECONDSUBROUND$

\noindent where

$firstSubRound \sqsubseteq  FIRSTSUBROUND$

$secondSubRound \sqsubseteq  SECONDSUBROUND$
\\

\begin{figure*}[htpb]
  \begin{center}
    \includegraphics [scale=0.8]
                      {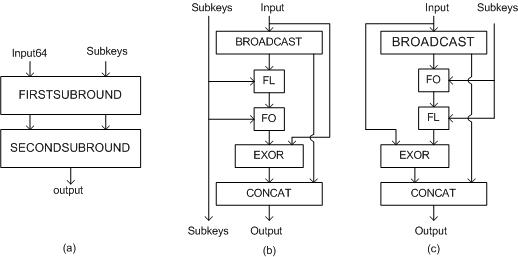}
    \caption{The processes (a) SINGLEROUND, (b) FIRSTSUBROUND, and (c) SECONDSUBROUND}
    \label{figs}
  \end{center}
\end{figure*}

In refining the function \textit{firstSubRound}, the datatypes could be refined as follows:
\\

$firstSubRound :: Int64 \rightarrow \lfloor [Int16] \rfloor_{6} \rightarrow (Int64, \lfloor
[Int16] \rfloor_{3})$
\\

Recalling the functional specification:

{\singlespace {\small \begin{verbatim}
 firstSubRound input64
  [kLo, kOo, kIo, kLe, kOe, kIe] =
  (l1++r1, [kLe, kOe, kIe])
  where
  [r1, r0]= [take 32 input64,
             drop 32 input64]

  [l1, t1, t2] = [(fullexor r0 t2),
                  (fL r1 kLo),
                  (fO t1 kOo kIo)]
\end{verbatim}}}

The process \textit{FIRSTSUBROUND} after getting its inputs, and
depending on the functional specification, firstly broadcasts the
input left half \textit{r1} to be used twice. Then, the subkeys
are produced to the processes \textit{FL} and \textit{FO} in the
order needed. The communications between \textit{FL} and
\textit{FO} is implicitly synchronised by the ($\parallel$)
operator. The output from \textit{FO} is passed to the process
\textit{EXOR} with the produced input right half. At this point,
the process \textit{CONCAT} is synchronising on the output of the
processes \textit{EXOR} and the broadcasted \textit{r1}. Finally,
the remaining subkeys are produced to be forwarded to the process
\textit{SECONDSUBROUND}. These processes are shown in
Figure~\ref{figs} (b).
\\

{\singlespace

\noindent $FIRSTSUBROUND =$

\noindent $(in_{1}?input64 \rightarrow SKIP) \interleave$

\noindent $(\interleave_{i=0,j=0}^{i=5,j=2}in_{2}.elements[i][j]? kss[i][j] \rightarrow SKIP);$

$BROADCAST_{2}(input64[32..63])[d/out] \parallel$

$((PRD(kss[0][0]) \parallel PRD(kss[0][1])) \rhd FL) \parallel$

$((PRD_{v}(kss[0]) \parallel PRD_{v}(kss[1])) \rhd FO) \parallel$

$(PRD(input[0..31]) \rhd EXOR))$

$\parallel CONCAT \parallel PRD_{v}(kss[3]) \parallel$

$PRD_{v}(kss[4]) \parallel PRD_{v}(kss[5])$

\noindent where

$fL \sqsubseteq FL$

$fO \sqsubseteq FO$ }
\\

Similarly, and for the function \textit{secondSubRound} the
refinement is done as follows:
\\

{\singlespace \noindent $secondSubRound ::(Int64, \lfloor [Int16] \rfloor_{3}) \rightarrow Int64$

\noindent $SECONDSUBROUND =$

$(in.fst?input64 \rightarrow SKIP)\interleave$

$(\interleave_{i=0}^{i=2}(\interleave_{j=0}^{j=2}in.snd.elements[i] ? kss[i][j]));$

$BROADCAST_{2}(input64[32..63])[d/out]\parallel$

$((PRD(kss[1]) \parallel PRD(kss[2])) \rhd FO) \parallel$

$((PRD(kss[0][0]) \parallel PRD(kss[0][1])) \rhd FL) \parallel$

$(PRD(input[0..31]) \rhd EXOR))\parallel CONCAT$}


\subsubsection{\normalsize\bf\quad\ Second Design}

In this design, the subkeys packs are passed in a stream setting to a single \textit{SINGLEROUND}
process. This stream refinement of \textit{foldl} implemented by \textit{SVFOLDL} will use the
\textit{SINGLEROUND} process to compute for the final desired folded result. This design affords
an economical use of computing resources. However, it is a quid pro quo for efficiency. This
\textit{CSP} network is pictured in Figure~\ref{KASUMI2D} and implemented as follows:
\\

$KASUMI = KEYSCHEDULE \parallel SVFOLDL(SINGLEROUND)$

\begin{figure*}[htpb]
  \begin{center}
    \includegraphics [scale=0.8]
                      {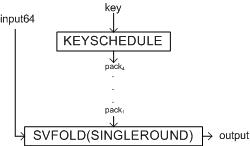}
    \caption{The process KASUMI, second design}
    \label{KASUMI2D}
  \end{center}
\end{figure*}


\subsubsection{Third and Fourth Designs}

The aim of introducing the third and fourth designs is to reduce the communication in the fine
levels, mainly inside the \textit{FL}, \textit{FI}, and \textit{FO} blocks. These blocks will be
implemented with basic operations instead of communicating processes. For example, an addition
will be implemented using a (+) operator instead of a process \textit{ADDITION}. The refinement of
the remaining blocks is to be the same. Also, the external communications with the \textit{FL},
\textit{FI}, and \textit{FO} blocks will be the same. The third design uses the new descriptions
for the F-blocks to modify the first fully-pipelined design, while the fourth design applies the
changes to the second stream-based design.


\section{\normalsize\bf\quad\ Reconfigurable Hardware
Implementations}\label{RHI}

\quad\ Based on the refined networks of \textit{CSP} processes we
include samples of the \textit{Handel-C} code used in the
realisation of the hardware circuit.

Getting a sample from \textit{KASUMI}'s main blocks, we present
the macro \textit{SingleRound} realising the processes
\textit{SingleRound}. The correspondence with the \textit{CSP}
description is very clear by refereing to the implementation
presented in the previous stage. In this macro, the macros
\textit{FirstSubRound} and \textit{SecondSubRound} are piped in
parallel to create the macro \textit{SingleRound} as follows:

{\singlespace {\small \begin{verbatim}
 macro proc SingleRound
    (input64, skeysVoV, output64) {

par{
FirstSubRound
    (input64, skeysVoV, midTuple);

SecondSubRound
    (midTuple, output64);}}
\end{verbatim}}}

The macros implementing the refined network of processes describing the \textit{KASUMI}, are
called from the macro \textit{Kasumi}. This macro implements the first design.

{\singlespace {\small \begin{verbatim}

macro proc Kasumi(input64,
 keysPacks, output64) {

 VFOLDL(input64, keysPacks,
    4, SingleRound, output64);}
\end{verbatim}}}


\section{\normalsize\bf\quad\ Performance Analysis and Evaluation}
\label{PAE}

\quad\ In this paper, we have demonstrated a methodology that can
produce intuitive, high-level specifications of algorithms in the
functional programming style. The development continues by
deriving efficient, parallel implementations described in
\textit{CSP} and realised using \textit{Handel-C} that can be
compiled into hardware on an \textit{FPGA}. We have provided a
concrete study that exploited both data and pipelined parallelism
and the combination of both. The implementation was achieved by
combining behavioural implementations 'off-the-shelf' of commonly
used components that refine the higher-order-functions which form
the building blocks of the starting functional specification.

The development is originated from a specification stage, whose
main key feature is its powerful \textbf{higher-level of
abstraction}. During the specification, the isolation from
parallel hardware implementation technicalities allowed for deep
concentration on the specification details. Whereby, for the most
part, the style of specification comes out in favor of using
higher-order functions. Two other inherent advantages for using
the functional paradigm are \textbf{clarity} and
\textbf{conciseness} of the specification. This was reflected
throughout all the presented studies. At this level of
development, the \textbf{correctness} of the specification is
insured by construction from the used correct building blocks. The
implementation of the formalised specification is tested under
\textit{Haskell} by performing random tests for every level of the
specification.

The correctness will be carried forward to the next stage of development by applying the provably
correct rules of refinement. The available pool of refinement formal rules enables a high degree
of \textbf{flexibility} in creating parallel designs. This includes the capacity to divide a
problem into completely independent parts that can be executed simultaneously (pleasantly
parallel). Conversely, in a nearly pleasantly parallel manner, the computations might require
results to be distributed, collected and combined in some way. Remember at this point, that the
refinement steps are \textbf{systematic} and done by combining off-the-shelf \textbf{reusable}
instances of basic building blocks.

In the following we will address the results found after
compiling, placing and routing, and running the proposed designs.
In Table~\ref{KTblEnSKeysSV} the key scheduling design occupied
8905 Slices and performed at a throughput of 27.7 Mbps. The
\textit{KASUMI} block algorithm in the stream-based second design
occupied 13225 Slices and performed at a throughput of 1.68 Mbps
(See Table~\ref{KTblEnc}). The third and fourth designs
outperformed the second design with speeds of 4.92 Mbps and 32
Mbps. The fourth design had a better running frequency (72.71 MHz)
than of the third design (49.06 MHz).

These testing results, as compared to the requirements and to
other hardware implementations, reveal the high cost of applying
the methodology in that manner. Even if some tuning were made,
tracking the critical paths in timing analysis to increase the
maximum possible frequency of the design does not promote an
elevated expectancy of the throughput. The high cost in hardware
resources arises from the applied systematic rules blinding
possibilities for intuitive ad hoc optimisations. The trials for
better speed could continue in a similar way to those undertaken
in the \textit{KASUMI} third and fourth designs. Nevertheless,
this lessens the use of communications on the fine-grained
processes levels.

\begin{table}
    \caption{Testing results of the key scheduling implementation}
    \label{KTblEnSKeysSV}
         \begin{center}
            \includegraphics [scale=0.8] 
                      {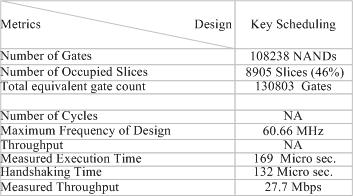}
         \end{center}
\end{table}

\begin{table*}[htpb]
    \caption{Testing results of the KASUMI block cipher implementation}
    \label{KTblEnc}
         \begin{center}
            \includegraphics [scale=0.7] 
                      {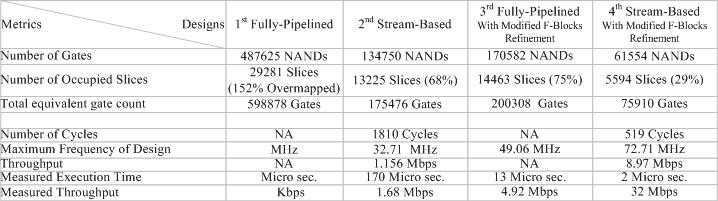}
         \end{center}
\end{table*}

\section{\normalsize\bf\quad\ Acknowledgement}

\quad\ I would like to thank Dr. Ali Abdallah, Prof. Mark Josephs, Prof. Wayne Luk, Dr. Sylvia
Jennings, and Dr. John Hawkins for their insightful comments on the research which is partly
presented in this paper.

\section{\normalsize\bf\quad\ Conclusion} \label{Con}

\quad\ Recent advances in the area of reconfigurable computing
came in the form of \textit{FPGAs} and their high-level
\textit{HDLs} such as \textit{Handel-C}. In this paper, we build
on these recent technological advances by presenting,
demonstrating and examining a systematic approach for synthesizing
parallel hardware implementations from functional specifications.
We have observed a case study from applied cryptography, namely
the \textit{KASUMI} algorithm for \textit{3GPP}. The testing of
the realised reconfigurable circuits allowed the ciphering with
\textit{KASUMI} in a throughput of 32 Mbps with an occupied area
of 5594 Slices. However, this confirms the conclusion showing the
expense of using the higher-level approach adopted. Future work
includes extending the theoretical pool of rules for refinement,
the investigation of automating the development processes, and the
optimisation of the realisation for more economical
implementations with higher throughput.


\section{\normalsize\bf\quad\ References}

 \makeatletter
    \def\thebibliography#1{\section*{\@mkboth
      {REFERENCES}{REFERENCES}}\list
     {[\arabic{enumi}]}{\settowidth\labelwidth{[#1]}\leftmargin\labelwidth
    \advance\leftmargin\labelsep
    \usecounter{enumi}}
    \def\newblock{\hskip .11em plus .33em minus .07em}
    \sloppy\clubpenalty4000\widowpenalty4000
    \sfcode`\.=1000\relax}
    \makeatother

{\singlespace
\bibliography{kasumi}
\bibliographystyle{unsrt}
}


\end{multicols}
\end{document}